\documentclass[%
 reprint,
superscriptaddress,
 amsmath,amssymb,
 aps,
prl,
]{revtex4-2}

\usepackage{graphicx, color}
\usepackage{dcolumn}
\usepackage{bm}
\usepackage[separate-uncertainty]{siunitx}
\usepackage[hidelinks]{hyperref}

\newcommand{\myvec}[1]{\boldsymbol{#1}}

\newcommand{\reffigpanel}[2]{Fig.~\ref{#1}#2}

\usepackage[draft]{changes}
\definechangesauthor[name={Zheng}, color=magenta]{zh}
\definechangesauthor[name={Michael}, color=olive]{mi}
\definechangesauthor[name={Simon}, color=orange]{si}
\definechangesauthor[name={Kris}, color=red]{kr}
\definechangesauthor[name={Matteo}, color=blue]{mt}

\newcommand{\ReplyAddition}[1]{{\color{black} #1}}
\newcommand{\ReplyAdditionPRR}[1]{{\color{black} #1}}

\begin{document}

\title{Colliding Pulse Injection of Polarized Electron Bunches in a Laser-Plasma Accelerator}

\author{S. Bohlen}
\affiliation{%
	Deutsches Elektronen-Synchrotron DESY, Notkestr. 85, 22607 Hamburg, Germany
}%
\author{Z. Gong}
\affiliation{%
	Max-Planck-Institut für Kernphysik, Saupfercheckweg 1, 69117 Heidelberg, Germany
}%
\author{M. J. Quin}
\affiliation{%
	Max-Planck-Institut für Kernphysik, Saupfercheckweg 1, 69117 Heidelberg, Germany
}%


\author{M. Tamburini}
\affiliation{%
	Max-Planck-Institut für Kernphysik, Saupfercheckweg 1, 69117 Heidelberg, Germany
}%
\author{K. P\~oder}%
\email{kristjan.poder@desy.de}
\affiliation{%
	Deutsches Elektronen-Synchrotron DESY, Notkestr. 85, 22607 Hamburg, Germany
}%

\date{\today}

\begin{abstract}
Highly polarized, multi-kiloampere-current electron bunches from compact laser-plasma accelerators are  desired for numerous applications. Current proposals to produce these beams suffer from intrinsic limitations to the reproducibility, charge, beam shape and final polarization degree. In this Letter, we propose colliding pulse injection as a technique for the generation of highly polarized electron bunches from pre-polarized plasma sources. 
Using particle-in-cell simulations, we show that colliding pulse injection enables trapping and precise control over electron spin evolution, resulting in the generation of high-current (multi-kA) electron bunches with high degrees of polarization (up to \SI{95}{\percent} for $>\SI{2}{kA}$). 
Bayesian optimization is employed to optimize the multidimensional parameter space associated with CPI to obtain percent-level energy spread, sub-micron normalized emittance electron bunches with \SI{90}{\percent} polarization using 100-TW class laser systems. 

\end{abstract}

\maketitle
Spin-polarized electron beams are of fundamental importance in atomic, nuclear and particle physics \cite{Tolhoek1956,Abe1995s,Anthony2004s,Schlimme2013s} due to the possibility of increasing the sensitivity of fundamental processes and enabling results competing with high-energy accelerators \cite{Androic2018s}. As such, polarized beams are also considered in many proposals for future colliders \cite{Shiltsev2021}, e.g., to enhance the searches at these machines for physics beyond the standard model \cite{Moortgat-Pick2008s}. Furthermore, polarized electrons can be used to generate polarized beams of photons \cite{Martin2012} or positrons \cite{Abbott2016s}, which are of great interest in material science applications \cite{Schultz1988}. 

Currently, the main sources of polarized electron beams are storage rings relying on radiative polarization (Sokolov-Ternov effect)~\cite{Sokolov1967} and polarized photocathodes~\cite{Pierce1975}. 
These machines rely on radio-frequency (RF) technology and are therefore large-scale and scarce. 
Laser-plasma acceleration (LPA)~\cite{Tajima1979, Esarey2009} offers an exciting alternative to conventional accelerator technology by utilizing 100GV/m-level acceleration gradients in plasma waves excited by an ultrahigh laser pulse in a tenuous plasma.
Alternative sources of polarized beams have been proposed~\cite{Batelaan1999, Dellweg2017}, including techniques to create spin-polarized electron beams with orders of magnitude higher peak currents than RF sources with compact LPAs. 
Concepts such as generation of polarized beams from pre-polarized plasma sources using density down-ramp injection (DDR) \cite{Wen2019}, self-injection (SI) \cite{Fan2022}, or Laguerre-Gaussian (LG) laser beams~\cite{Wu2019} have been put forward. Several proposals also exist for employing beam-driven plasma accelerators~\cite{Wu2019PBA,Nie2021,Nie2022} or interactions with ultrahigh-intensity laser pulses~\cite{delSorbo2017, Li2019, Seipt2019, Xue2022, Li2022} to generate spin-polarized electron bunches.

Among these novel polarized electron beam source proposals, LPA-based concepts show the highest near-term potential. Yet all currently proposed methods suffer from limitations hindering their practical implementation.
The azimuthal magnetic field inside the LPA acceleration cavity decreases the polarization of electrons injected off-axis in the DDR case, limiting the driver laser to $a_0 \lesssim 1$ and consequently the charge of highly polarized electron bunches to 100s~fC~\cite{Wen2019}.
Here, $a_0\simeq0.85\lambda_0[\SI{}{\micro\metre}] \sqrt{I_0[10^{18}\mathrm{Wcm^{-2}}]}$ is the peak normalized laser vector potential, where $\lambda_0$ and $I_0$ are the laser wavelength and peak intensity, respectively.
Furthermore, the low $a_0$ necessitates very sharp density transitions, which are difficult to generate. 
Studies have shown that highly spin-polarized beams can be self-injected, but deviations from perfect spherical symmetry will severely downgrade beam polarization~\cite{Fan2022}. And while high current and polarization can be achieved using LG laser drivers, the resulting annular electron beams can be impractical and very sharp density transitions are again required~\cite{Wu2019}. 

In this Letter, we propose using colliding pulse injection (CPI)~\cite{Umstadter1996,Esarey1997,Faure2006} to create highly polarized ($\SI{>90}{\percent}$), high charge (tens of pC) and low normalized emittance (\SI{<1}{\milli\metre\milli\radian}) electron beams with percent-level energy spread from a pre-polarized plasma source. The easily adjustable degrees of freedom stemming from the colliding laser pulse properties enable control over the phase-space volume of the injected bunch and spin depolarization during the injection process.
We employ Bayesian optimization~\cite{Shalloo2020s, Jalas2021} to tune some of the available degrees of freedom to demonstrate generation of high-quality, highly polarized electron bunches, thereby exemplifying the huge potential of this technique.

\begin{figure}[t]
    \centering        
    \includegraphics[width=\columnwidth]{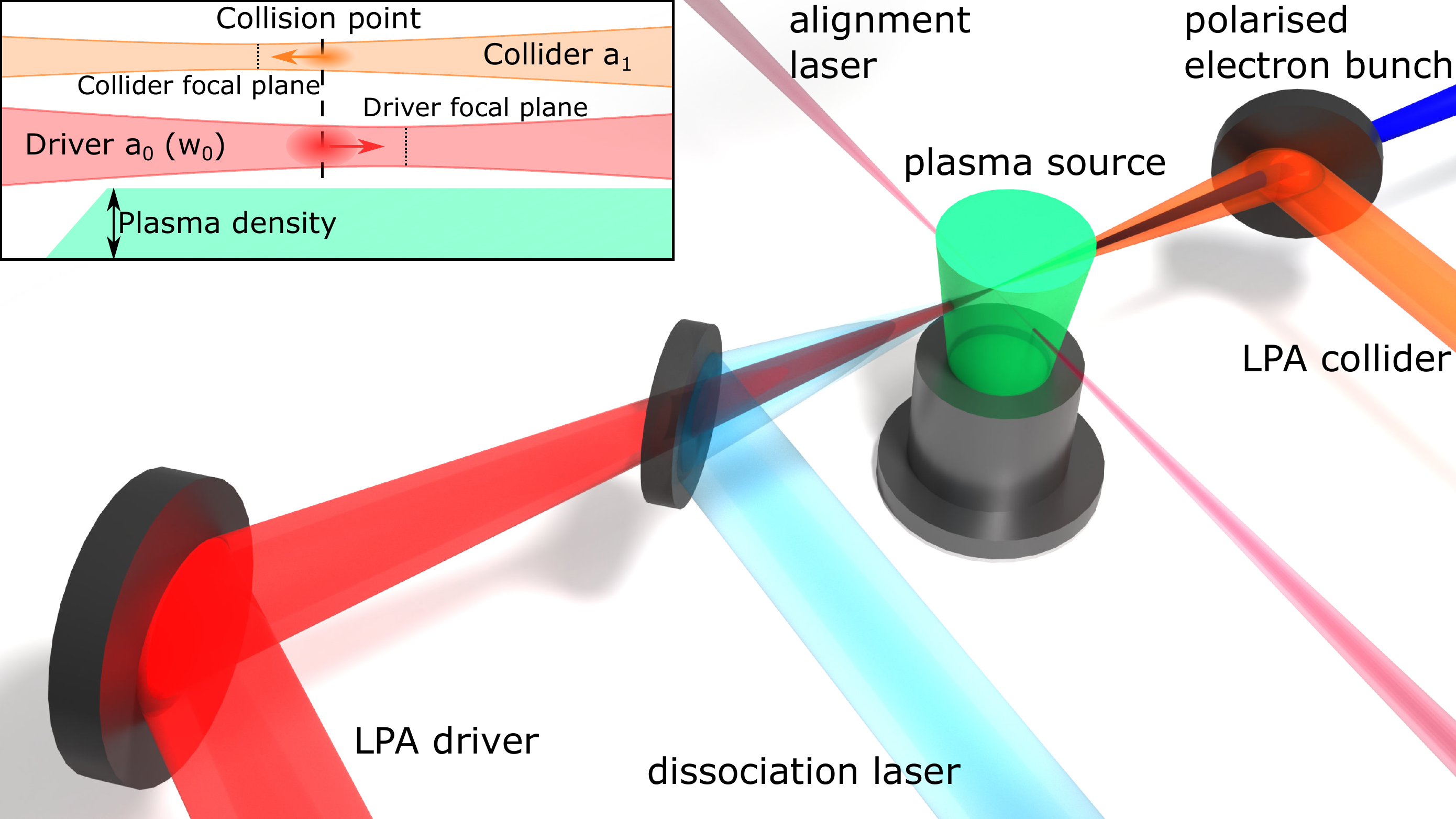}
    \caption[Experimental Setup]{Schematic setup of a laser plasma accelerator employing CPI for the generation of spin-polarized electron beams from a pre-polarized plasma source; the different laser beams are discussed in the main text. 
    The inset shows parameters that can be varied to optimize the injected electron bunch parameters: plasma density $n_e$, collision point $z_c$, focal plane $z_{f, 0}$ ($z_{f, 1}$) and intensity $a_0$ ($a_1$) of the driver (collider) laser.}
    \label{Fig:ExpSetup}
\end{figure}

A schematic setup for polarized LPA using CPI 
is shown in Fig.~\ref{Fig:ExpSetup}.
CPI employs a second laser pulse (orange) counter-propagating to the LPA driver (red), with the standing wave set up in the vicinity of their collision point enabling trapping of background electrons.
\ReplyAddition{The plasma source could be pre-polarized using hydrogen halide molecular dissociation~\cite{Rakitzis2003,Rakitzis2004,Sofikitis2008,Sofikitis2017,Sofikitis2018,Hutzen2019}. A sub-nanosecond alignment laser 
perpendicular to the LPA driver (purple) 
aligns the molecular bonds. Subsequently a circularly polarized UV pulse (teal) dissociates the halide molecule}\ReplyAdditionPRR{, resulting in two polarized valence electrons from the hydrogen-halide bond. After full ionization of outer shell electrons this would result in a maximum polarization of 25\% for hydrogen halides.} \ReplyAddition{The delay from dissociation to injection into the wakefield must be of the order of 10s~ps, much smaller than the time for hyperfine coupling transferring polarization from electrons to nucleus~\cite{Sofikitis2017}.} 
\ReplyAdditionPRR{To increase the pre-polarisation degree towards 100~\%,} the dissociated hydrogen atoms with their spin-polarized electrons could be spatially separated from the halide atoms via resonance-enhanced multiphoton ionization~\cite{Wu2019, Spiliotis2021} \ReplyAddition{or other methods~\cite{RakitzisPC}.}
\ReplyAdditionPRR{Alternatively}, the pre-polarization technique could \ReplyAddition{potentially} be extended to pure hydrogen \ReplyAdditionPRR{in the future}, e.g. using lasers with $\lambda\lesssim\SI{100}{nm}$ for the dissociation, which are only recently becoming available~\cite{Drescher2021}\ReplyAddition{, or other developments that enable the production of spin polarized hydrogen.}
\ReplyAddition{In an experiment, only the restricted volume where electrons are trapped must be pre-polarized.}
CPI is greatly advantageous as the location of the collision point can easily be adjusted to this pre-polarized volume; 
\ReplyAddition{indeed micron-scale overlap of multiple laser pulses at arbitrary positions in a plasma has already been demonstrated~\cite{Bohlen2022}.}
In this Letter, we assume a \SI{100}{\percent} pre-polarized plasma source to study the achievable spin-polarization using CPI. 

\begin{figure}[b]
	\centering        
	\includegraphics[width=\columnwidth]{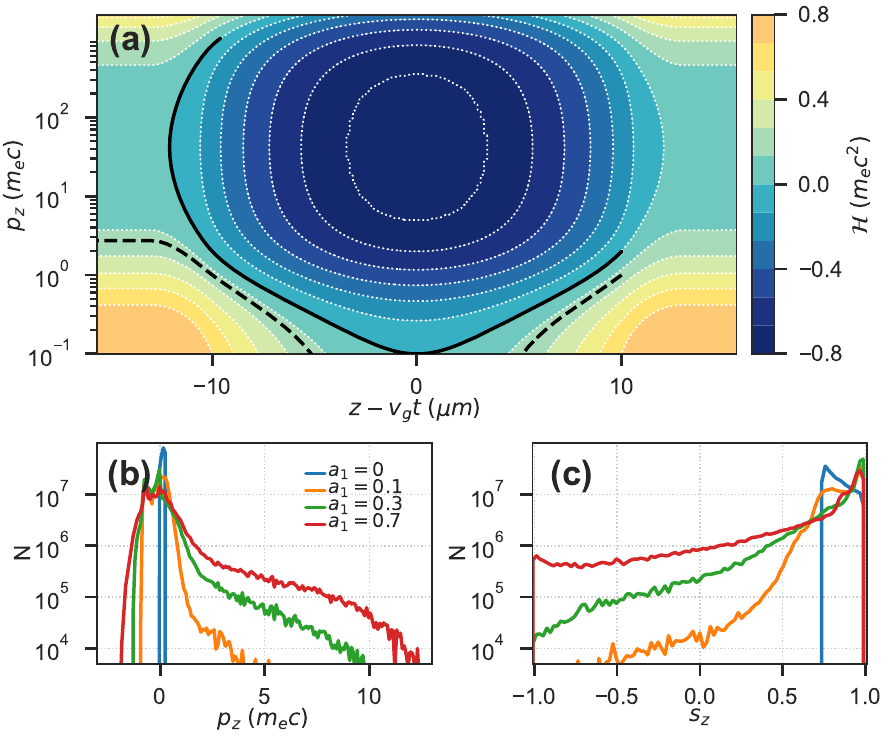}
	\caption[Theory]{\textbf{(a)} The Hamiltonian of a test electron in a plasma wave, showing an untrapped (dashed line) and trapped (solid line) orbit. \textbf{(b)} Longitudinal momentum distribution of test electrons with $r<w_1$ after interacting with collider pulse of varying intensity in vacuum, showing onset and increase of stochastic electron heating. \textbf{(c)} Histogram of $s_z$ after interaction with a colliding pulse, highlighting the reduction of polarization due to stochastic heating. Laser parameters were $a_0=2.5$, $w_0=\SI{20}{\micro\metre}$, $w_1=\SI{5}{\micro\metre}$.}
	\label{Fig:Theory}
\end{figure}

Bunch injection in CPI can be described in terms of electron trapping, utilizing the Hamiltonian $\mathcal{H}(p_z, \xi)$ inside the potential well of a non-linear plasma wave~\cite{Esarey1995, Gong2023}, where $\mathcal{H}=\mathrm{const}$ along an electron orbit, as shown in \reffigpanel{Fig:Theory}{a}. 
An electron in an \textit{untrapped} orbit (black dashed line in \reffigpanel{Fig:Theory}{a}), where it simply passes the plasma wave, can move to a \textit{trapped} orbit (black solid line in \reffigpanel{Fig:Theory}{a}) where it is accelerated inside the plasma wave by gaining momentum. The interference of driver and collider laser with peak normalized vector potentials $a_0>a_1$, respectively, creates a standing ($v_{ph}=0$) beat wave causing stochastic heating of electrons~\cite{Mendonca1983, Zhang2003, Bourdier2005, Malka2009}. The longitudinal momentum $p_z$ gained in the heating process enables some electrons to become trapped in the plasma wave~\cite{Umstadter1996, Esarey1997, Faure2006, Malka2009}. In \reffigpanel{Fig:Theory}{b} the longitudinal momentum distribution of electrons is shown for varying colliding pulse strengths, showing how increasing $a_1$ causes more heating, resulting in a larger fraction of electrons gaining higher amounts of $p_z$. The theoretical framework of spin-polarized bunch injection in CPI is developed in more detail in Ref.~\onlinecite{Gong2023}.


To study the electron spin evolution in CPI, 
the Thomas-Bargmann-Michel-Telegdi equation~\cite{TBMT} describing spin dynamics of electrons can be used. 
The spin vector $\myvec{s}$ of unit length evolves according to classical spin dynamics in time-varying electric and magnetic fields as
\begin{equation}
    \frac{\mathrm{d}\myvec{s}}{\mathrm{d}t} = (\myvec{\Omega_T} + \myvec{\Omega_a}) \times \myvec{s},
    \label{Eq:TMBT}
\end{equation}
with
\begin{subequations}
\label{Eq:omega}
\begin{eqnarray}
    \myvec{\Omega_T} &=& \frac{q_e}{m_e}\left(\frac{1}{\gamma_e}\myvec{B} - \frac{\myvec{\beta}}{1+\gamma_e}\times\frac{\myvec{E}}{c} \right), \\
    \myvec{\Omega_a} &=& a_e \frac{q_e}{m_e}\left[\myvec{B} - \frac{\gamma_e}{1+\gamma_e}\myvec{\beta}(\myvec{\beta}\cdot\myvec{B}) - \myvec{\beta}\times\frac{\myvec{E}}{c} \right],
\end{eqnarray}
\end{subequations}
where $\gamma_e =1 / \sqrt{1 - \beta^2}$ is the Lorentz factor of the electron, $\myvec{\beta}=\myvec{v}/c$ is the normalized velocity and $m_e$, $-q_e$, $a_e \approx 1.16\times10^{-3}$ are the electron mass, charge and anomalous magnetic moment, respectively. 
Given that the spin precession rate depends on $\myvec{\beta}$, it is clear that the chaotic motion due to stochastic heating in the standing beat wave also results in stochastic evolution of the spin-vector $\myvec{s}$. 
This causes a fraction of initially polarized (e.g. along $\hat{\myvec{e}}_z$) electrons to become depolarized, as shown in \reffigpanel{Fig:Theory}{c}, highlighting how the stochastic heating from increasing $a_1$ results in a larger amount of depolarized electrons. 
\ReplyAddition{Unlike in the previously published regimes~\cite{Wen2017, Fan2022, Wu2019}, where the spins precess due to the strong azimuthal fields on the periphery of the bubble, in CPI the depolarisation is a result of stochastic effects,} as discussed in-depth in another publication~\cite{Gong2023}.
\ReplyAddition{In CPI, the stochastic spin depolarization of the injected electrons can be suppressed by varying $a_1$ and $w_1$, while injecting close to the axis inherently avoids strongly depolarizing azimuthal B-fields;} $z_c$ can be varied to optimize energy gain and energy spread of the bunch. These additional degrees of freedom enable CPI to be used to generate highly polarized, high-quality electron bunches from pre-polarized plasma targets. 

We used the quasi-3D particle-in-cell code FBPIC~\cite{Lehe2016,Jalas2017} to study the generation of polarized electron beams using CPI. The code was modified to include particles' spin properties~\cite{Gong2023} by implementing Equation~\ref{Eq:TMBT}. Other spin effects were not modeled~\cite{Mane2005} as the Stern-Gerlach force is orders of magnitude smaller than the Lorentz force~\cite{Wen2017} and the timescale for Sokolov-Ternov effect~\cite{Sokolov1967}, even in the hundreds of GV/m field strengths present in LPAs, is on the order of microseconds, being much longer than the typical nanosecond duration of laser-plasma acceleration~\cite{Thomas2020}.
Scaling of the electron bunch spin-polarization in CPI was studied by performing a 2D grid scan varying the focal spot size $w_1$ and $a_1$ of the colliding laser. Both lasers had a FWHM pulse duration of \SI{30}{fs} and $\lambda=\SI{800}{nm}$. The plasma density profile in the simulations consisted of a \SI{50}{\um} linear ramp followed by a plateau with $n_e=\SI{1e18}{cm^{-3}}$. The simulation box had a length of \SI{70}{\um} and a radius of \SI{60}{\um} with 3000 and 500 cells, respectively, and moved in the positive $z$ direction at the speed of light. The number of particles per cell was set to 2, 2, 4 for the longitudinal, radial and azimuthal directions, respectively, with three azimuthal modes used. The colliding laser was injected backwards from an antenna 
and was set to collide with the driver laser at $z=\SI{100}{\micro\metre}$, which was also the focal plane for both the driver and collider laser. A fully pre-polarized and pre-ionized plasma was modeled, with the initial spin vector of the background electrons set to $\myvec{s} = \myvec{\hat{e}_z}$. The electron beam parameters were analyzed after propagating the driver beam through \SI{500}{\micro\metre} of plasma, with the polarization calculated as $P=|\Sigma_i^N \myvec{s}_i/N|$.

\begin{figure}[t]
	\centering        
	\includegraphics[width=\columnwidth]{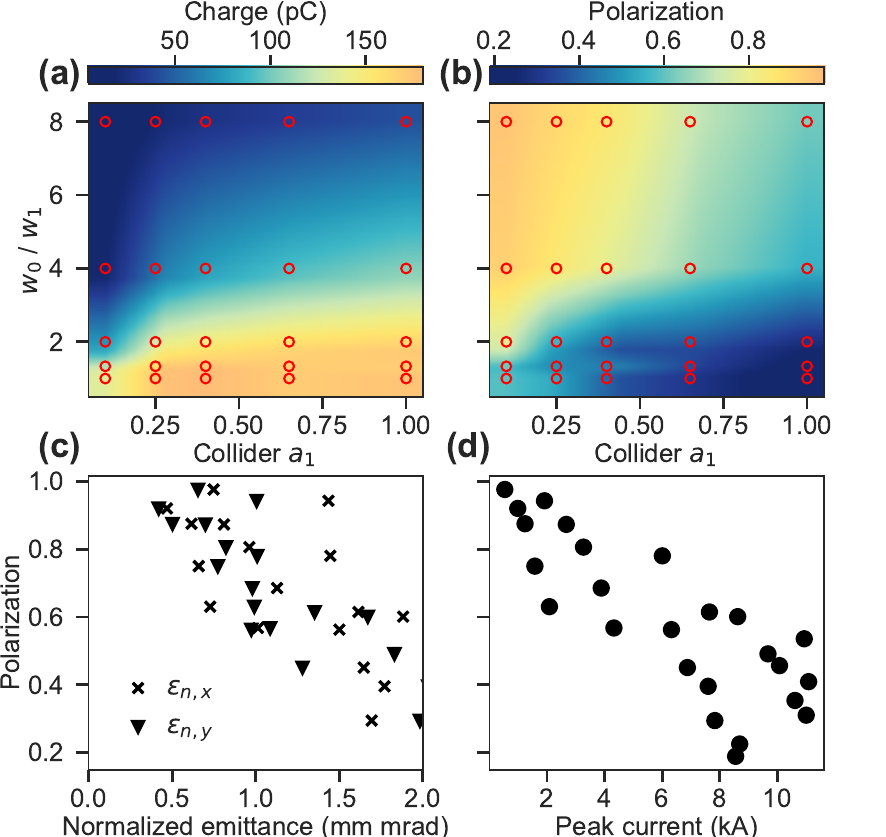}
	\caption[Gridscan]{Variation of spin-polarized electron beam parameters. \textbf{(a)} Charge and \textbf{(b)} polarization as a function of $a_1$ and $w_1$ with $w_0=\SI{20}{\micro\metre}$ and $a_0=2.5$. The red circles denote individual simulation results while the color scale is an interpolation over these points. \textbf{(c)} The normalized emittance and polarization of the generated beams. \textbf{(d)} Peak current and polarization variation. 
	}
	\label{Fig:Scanresults}
\end{figure}

Results of this scan are depicted in Fig. \ref{Fig:Scanresults}, highlighting that even without careful optimization, CPI allows highly polarized bunches with sub-micron normalized emittance to be generated. Panels (a) and (b) in Fig.~\ref{Fig:Scanresults} show that larger $w_1$ as well as increased $a_1$ generate bunches with higher charge at the expense of spin-polarization. This occurs as both a larger focal spot size $w_1$ and higher intensity $a_1$ lead to a larger volume of background electrons being stochastically heated, leading to a larger fraction of electrons to become substantially depolarized.
Thus, highly polarized beams $P\SI{>70}{\percent}$ are generated employing moderate collider laser intensities $a_1 \leq 0.5$ and small collider spot sizes relative to the driver laser. 
\reffigpanel{Fig:Scanresults}{c} shows that normalized emittance on the order of \SI{1}{mm~mrad} can be expected, especially for beams with high polarization in the range of \SIrange{60}{100}{\percent}. The bunch charge is up to \SI{50}{pC} for beams with polarization as high as \SI{90}{\percent};  correspondingly, peak currents of several kiloamperes are reached, as depicted in \reffigpanel{Fig:Scanresults}{d}. 

These results vastly outperform previously proposed schemes for LPA-generated polarized beams, both for the case of DDR~\cite{Wen2019, Wu2019} and self-injection~\cite{Fan2022}, particularly in terms of the generated bunch charge. We also note that in Ref.~\onlinecite{Wu2019}, the longitudinal polarization of a gaussian electron beam was calculated to vary as $P=\mathrm{sinc} (0.0072 I_{peak})$, meaning a fully depolarized beam $P=0$ would be expected for $I_{peak}=\SI{4.5}{kA}$. This limitation is, however, derived from and inherently linked to the density downramp profile used in that study. From our work it is clear that highly polarized, high-current electron bunches can indeed be generated using gaussian laser pulses.

The use of the colliding laser pulse to inject electrons introduces several free parameters that can be tuned to generate electron bunches with desired application-specific parameters. 
Together with the driver laser and plasma properties, this spans a large, multi-dimensional parameter space. 
Recent work has shown that Bayesian optimization (BO) algorithms are very well suited for finding the optimal parameters for such multi-dimensional problems \cite{Shalloo2020s,Jalas2021}. We used the BO toolkit \textsc{optimas}~\cite{FerranPousa2022, FerranPousa2023} to optimize a set of easily adjustable input parameters to generate electron bunches with high charge, low energy spread and high longitudinal polarization. In these FBPIC simulations, the plasma profile consisted of a \SI{100}{\micro\metre} long linear ramp starting at $z=0$ followed by a density plateau with $n_e$. 
The varying (\textit{to be optimized}) parameters (c.f. Fig.~\ref{Fig:ExpSetup}) were the collider intensity $a_1$, the collision point of the laser pulses $z_c$, the focal plane $z_f$ (overlapping for both lasers, i.e. $z_f=z_{f, 0}=z_{f, 1}$), the plasma density $n_e$ and the driver intensity $a_0$. A fixed driver laser power $P_0=\SI{100}{TW}$ was used, thus the variation of $a_0$ also changed $w_0$ as $a_0^2 w_0^2=\mathrm{const}$. The collider spot size was fixed at $w_1=\SI{5}{\micro\metre}$ and both pulses had a FWHM duration of \SI{30}{fs}. The simulation box length was \SI{60}{\micro\metre} with 3000 cells in the longitudinal direction, whereas the radius of the box was scaled to always be $r>2.5w_0$, with $\Delta r=\SI{120}{nm}$. We note that in this optimization the key free experimental parameters were varied, i.e. these parameters would also be tuned in case of manual optimization.

\begin{table}[t]
	\begin{ruledtabular}
		\begin{tabular}{l l l}
		Beam parameter & Value & Unit \\ \hline
	
		Mean energy               & 85.2 & MeV     \\
		Energy spread (rms)  & 4.4 & \%              \\
		Peak current                & 3.6 & kA       \\
		Bunch duration (rms) & 3.8 & fs       \\
		Charge                         & 31.8 & pC      \\
		Normalized emittance, $x$-plane             & 0.90 & mm mrad \\
		Normalized emittance, $y$-plane             & 0.84 & mm mrad \\
		Spin polarization        & 0.90 & \\
	
		\end{tabular}
	\end{ruledtabular}
	\caption{Properties of the bunch with lowest value of $f$ achieved by BO of the polarized CPI scheme.}
	\label{Tab:Optresults}
\end{table}

The beam properties were analyzed within the plasma after propagating the driver laser pulse for \SI{1}{mm}. The fitness function was set to $f=-\sqrt{QE_m}/[ \Delta E (1-P)]$, where $Q$ is the beam charge, $E_m$ is the median energy, $\Delta E$ is the median absolute deviation of the energy and $P$ is the beam polarization. The BO was run for 100 iterations. Despite the use of only a limited set of varied parameters, the BO quickly converged on a parameter region that results in generation of high-quality electron beams. The parameters for reaching the lowest $f$ were $a_0=2.68$, $w_0=\SI{20.4}{\micro\metre}$, $z_f=\SI{135.3}{\micro\metre}$, $z_c=\SI{281.4}{\micro\metre}$ and $n_e=\SI{8.2e17}{\per\centi\metre\cubed}$.
The properties of the bunch resulting in lowest value of $f$ are presented in Tab.~\ref{Tab:Optresults}, showing that electron bunches with few percent energy spread, tens of picocoulomb of charge and polarization of 90\% are achievable given the reasonable laser parameters employed. 

\begin{figure}[t]
    \centering        
    \includegraphics[width=\columnwidth]{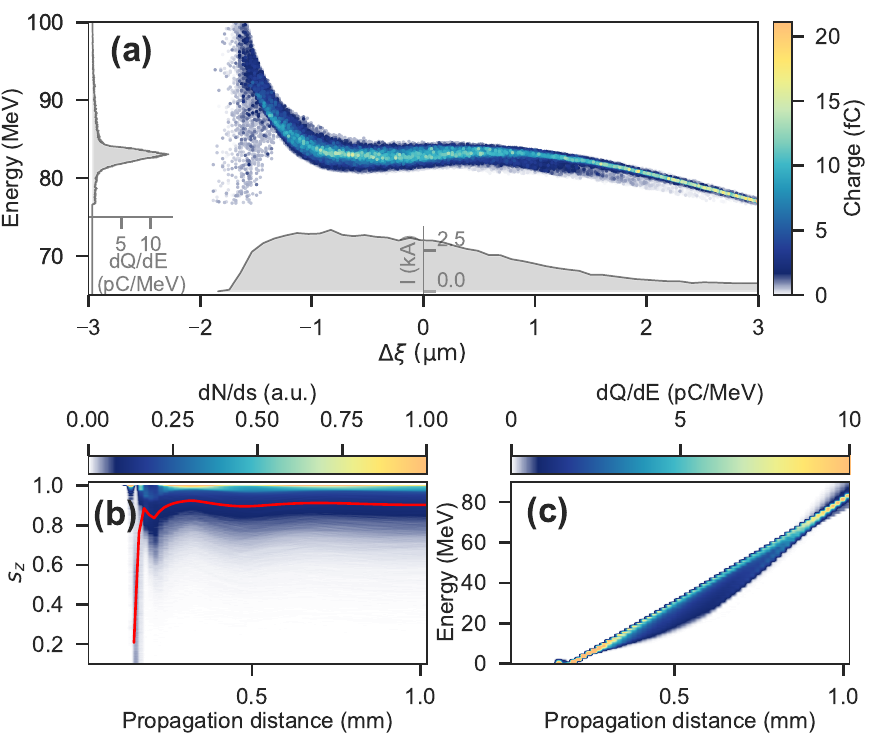}
    \caption[Experimental Setup]{Optimized highly polarized electron beam. \textbf{(a)} Longitudinal phase-space of the beam. \textbf{(b)} Evolution of $s_z$ during acceleration. The color scale shows distribution of $s_z$ while the red line is $P=\left< s_z \right> $. \textbf{(c)} Evolution of beam energy.}
    \label{Fig:Optresults}
\end{figure}

The longitudinal phase space of the optimized beam is shown in \reffigpanel{Fig:Optresults}{a}, highlighting the flattened energy distribution. The evolution of the longitudinal spin component of the accelerated electrons depicted in Fig.~\ref{Fig:Optresults}b indicates that the spin distribution and beam polarization is stable at  $\sim\SI{90}{\percent}$ after the electrons become highly relativistic, as has already been shown previously \cite{Wen2019, Vieira2011}. Consequently, the electron bunch energy could be increased by extending the acceleration length without significantly altering the beam polarization.
\ReplyAddition{As seen from Eq.~\ref{Eq:TMBT}, the spin precession depends only on applied fields, which suggests lack of additional depolarisation in plasma outcoupling downramp. This was confirmed by simulations of beam extraction from the plasma using emittance-conserving plasma downramps~\cite{Dornmair2015}}. 
The electron energy spectrum evolution depicted in Fig.~\ref{Fig:Optresults}c shows that the bunch is almost optimally beam-loaded~\cite{Kirchen2021}, while also indicating that dephasing or depletion are not yet limiting the energy gain. 100-TW class lasers have already been used to demonstrate multi-GeV energy gains~\cite{Clayton2010, Mirzaie2015}, suggesting the electron beam energy could be extended to GeV and beyond in a sufficiently long plasma source. 

We note that the optimization presented above can be further improved upon in the future, e.g. by allowing  $w_1$ or the focal plane of the collider laser to be changed. Additionally, the transverse or temporal profile of the collider laser could be varied to create tailored bunch spatial or current profiles. For example, a radially flattened beat-wave pattern arising from an annular collider pulse could potentially further increase the injected charge while maintaining the high polarization levels demonstrated in this work. 
Overall, CPI shows great potential to produce high-quality and high-current polarized electron beams required for a host of applications. 

\ReplyAddition{The results above assume a fully pre-polarized plasma $\bar{s}_z=1$, which may be difficult to achieve experimentally. Simulations using optimised parameters leading to beams depicted in Fig.~\ref{Fig:Optresults} were carried out to study the final beam polarisation $P$ dependence on the initial plasma pre-polarisation $\bar{s}_z$. A constant depolarisation of 0.9 was found for $\bar{s}_z=[0.3,1]$. Therefore, our results demonstrating low depolarisation will be instrumental in near-term experimental demonstrations for spin-polarized LPA, even if performed with lower initial pre-polarization.}

In summary, we show that narrow energy spread, low emittance and highly polarized electron bunches can be generated using CPI in a pre-polarized plasma source.
The additional degrees of freedom in CPI enable precise control of the injection process and subsequent beamloading, all the while allowing high spin-polarization of the bunch to be retained. Our work represents orders of magnitude improvements of the delivered beam charge while retaining high polarization upon previous results, bringing the advent of high-current polarized LPAs to within reach of currently commissioned laser systems.
A CPI-based, compact, thoroughly tunable source of highly polarized electron beams is a very promising candidate for injectors at future storage rings or linear collider setups or as an injector upgrade for existing polarized storage rings~\cite{Hillert2017s}. Further, LPA employing polarized CPI represents a long-awaited tool to expand access to scientific and industrial applications requiring polarized particle beams.

\begin{acknowledgments}
The authors acknowledge funding from the DESY Strategy Fund.  This work was supported by the Maxwell computational resources at DESY. The authors gratefully acknowledge the Gauss Centre for Supercomputing e.V. (www.gauss-centre.eu) for funding this project by providing computing time through the John von Neumann Institute for Computing (NIC) on the GCS Supercomputer JUWELS at Jülich Supercomputing Centre (JSC).
\end{acknowledgments}


%

\end{document}